# Additive Manufacturing of Fibrous Sound Absorbers

William Johnston[f] and Bhisham Sharma[l]

Department of Aerospace Engineering, 1845 Fairmount Street, Wichita State University, Wichita, KS 67260

[f]First author: E-mail: wjjohnston@shockers.wichita.edu; Tele: +1-316-978-3199

[l]Corresponding author: E-mail: bhisham.sharma@wichita.edu; Tele: +1-316-978-3199

**Abstract**

We investigate the possibility of additively manufacturing fibrous sound absorbers using fused deposition modeling. Two methods for 3D printing fibers are proposed. The fiber bridging method involves extruding the filament between two points with no underlying supports. The extrude-and-pull method involves extruding a filament droplet before pulling away the extruder rapidly to generate thin fibers. Both methods can produce fibers with aspect ratios greater than 100. Optical microscopy is used to investigate the effect of various printing parameters on the fiber characteristics. The sound absorption coefficient of samples printed using the two techniques are measured using a two-microphone normal incidence impedance tube setup. Effects of printing parameters and fiber density variables are experimentally studied. The experimental studies are supported by the Johnson-Champoux-Allard semi-empirical analytical model informed using an inverse characterization approach. The analytical model is then utilized to understand the effect of fiber parameters on the acoustical transport parameters. It is observed that the two methods result in individual fibers with distinct characteristics. On average, the fiber bridging method results in thicker fibers, which results in comparatively higher sound absorption. However, the extrude-and-pull method results in fibers with hair-like characteristics (thick base with progressively decreasing thickness) and one may easily incorporate it within existing additive manufacturing routines to add fibers to a base surface, thus opening up a new route towards fiber-enhanced multifunctional structures.

## 1. Introduction

Absorbing unwanted sound waves without excessively constraining the air flow path is a recurring challenge across various engineering fields. In aeronautics, this challenge manifests in the design of absorptive acoustic liners required to absorb aircraft engine noise and help meet Federal Aviation Administration's increasingly stringent noise certification standards—while the liner must absorb the turbofan-generated noise, it must not restrict the aerodynamic flow path and reduce engine efficiency [1]. Computer server designers face a similar challenge [2]: the noise emanated from internal cooling fans must be reduced without blocking the flow path and reducing cooling performance. Similar examples are prevalent in fields ranging from automotive engineering [3] to automatic hand dryer manufacturing [4].

Porous materials are the most commonly employed solution to this ubiquitous problem. Depending on the structure of their solid phase, porous materials may be classified as cellular, fibrous, or granular [5]. Regardless, they all fall into the class of resistive sound absorbers—the dominant absorption mechanism is the dissipation of incident acoustic energy through visco-inertial and thermal losses [6]. While this mechanism limits their absorption performance at low frequencies— frequencies below 1000 Hz may be considered low for most applications—they provide attractive absorption values at medium-to-high frequencies [6]. This combined with their low weight, low cost, and easy availability, makes them an attractive sound absorption solution for applications typically requiring noise mitigation between the 1000–6000 Hz frequency range. While thicker layers can improve their low frequency absorption—the sound absorption coefficient is a not an inherent material property and is thickness-dependent—volume restrictions often make this infeasible. At lower frequency ranges, reactive sound absorbers such as Helmholtz resonators [7, 8] or stiffened membranes with [9, 10] or without [11] embedded masses provide significantly better absorption behavior, though their performance is often limited to a narrow frequency range. Recently, inspired by the idea of locally resonant elastic and acoustic metamaterials [10, 12, 13], some researchers have proposed the idea of structures that combine reactive elements within resistive porous materials to help improve their low-frequency absorption performance [14-17]. Such "co-dynamic" structures [14] offer an alternative route towards better sound absorption devices, though fabricating them using traditional methods at scales and rates necessary for wider adoption would be a challenge.

Additive manufacturing stands to play an important role in the development of improved porous sound absorbers. While traditional methods for fabricating foams [18] and fibers [19, 20] are well-established and allow fabrication at rates necessary to keep costs down, they offer limited ability to tailor the porous microstructures and are restricted to specific material choices. Additive manufacturing methods have opened up new possibilities for microstructural topologies, which were previously infeasible; microstructures that help improve sound absorption behavior while enabling multifunctionality [21] and the creation of materials that simultaneously provide improved stiffness [22], thermal [23], electromagnetic shielding [24], and energy absorption properties [25], among others. Consequently, researchers have recently begun investigating the potential of 3D printing porous sound absorbers. Liu et al. [26] studied the acoustic properties of porous absorbers with cylindrical through-holes printed with polycarbonate using a stereolithographic (SLA) method. For 10 mm thick samples with 0.8 mm pore diameters, they obtained a high absorption peak between 2000 and 4000 Hz. Though the high absorption region is relatively narrow, they showed that varying the pore angle can help lower the peak absorption frequency. Changing the angle from 0° to 45° also reduced the absorption values. Later, the same authors investigated the absorption performance of multilayered micro-perforated panel absorbers with a 3D printed front layer backed by an unspecified porous material [27] and showed that the multilayered system provides significantly higher absorption than the porous material itself. Guild et al. [28] used the fused deposition modeling (FDM) printing method to fabricate simple cubic lattice absorbers with a flexible thermoplastic polyurethane (TPU) material. The choice of TPU allowed them to print a flexural element as the absorber backing, which helped increase the absorption performance over the measured frequency range. Inspired by cereal straws, Huang et al. [29] leveraged the FDM printing method to study the acoustic properties of an idealized packed straw bale structure and showed that the observed high-absorption behavior results from a combination of visco-thermal diffusion and quarter-wavelength resonances within straws. Along similar lines, Koch [30] took inspiration from natural reeds and demonstrated a 3D printed design for aircraft noise reduction applications. Wojciechowski et al. [31] and Yang et al. [32] used SLA printing to fabricate and study the acoustic behavior of porous structures with various triply periodic minimal surfaces. Of the studied surfaces, the diamond topology provided the highest absorption. Further, Wojciechowski et al. also showed that the absorption performance can be improved using a stepwise porosity gradient and by tailoring the pore angles. Interestingly, converse to observations by Liu et al., Wojciechowski et al. report that non-zero angles with respect to the incident waves result in higher absorption without a significant shift in the absorption peak frequency. Yang et al. [33] investigated the acoustical performance of multilayered, micro-perforated panels (MPPs) printed using selective laser sintering and showed that they provide wider absorption bands than conventional MPPs. Recently, Fotsing et al. [34] studied the acoustic properties of samples with micro-rods printed using FDM printing and demonstrated that such materials offer a broadband sound absorption behavior which can be adjusted by varying their lattice parameters. For the same structural configuration, Boulvert et al. [35] studied the effect of fabrication defects inherent in the FDM process and concluded that not accounting for such defects in the modeling procedure may lead to underestimation of their absorption coefficient.

An underlying thread of all research conducted so far is that the focus has exclusively been on fabrication of porous foams. Can additive manufacturing also be used to fabricate fibrous sound absorbers? Here, borrowing from the field of textile engineering [36], we define a fiber as an individual threadlike structure with an average length-to-thickness ratio equal to or greater than 100. The sound absorption properties of fibrous structures have been widely studied [19, 20]; however, to the best of our knowledge, no attempts have so far been made to additively manufacture fibrous absorbers.

In this paper, we explore the fabrication of fibrous structures using FDM printing within the context of their use for sound absorption applications. A typical FDM workflow involves using a 3D modeling software to generate requisite CAD models, converting these models from their native format to a stereolithographic (STL) file format, and then translating the model information into machine-readable commands using a 'slicing' software. The slicing software decomposes (or slices) the 3D model into individual layers and generates a corresponding G-code—a machine-readable script containing the tool-path information required by the FDM printer to print the object. The G-code directly controls the print process and contains information such as the model's spatial coordinates on the print bed, filament extrusion volume and rate, nozzle temperature, and the extruder travel speed while printing each layer. Here we present two different approaches for printing fibers using the FDM process: Fiber bridging—a technique that relies on the bridging capabilities of the printer; and Extrude-and-Pull—a technique, originally proposed by Laput et al. [37], that involves altering the G-code. Our results indicate that both methods result in fibers with different thickness characteristics, which can be tailored by altering the printing parameters. The absorption behavior of the printed fibrous absorber samples is well represented by the commonly used Johnson-Champoux-Allard (JCA) model [6, 38, 39]. Further, we use this model to gain insights into the relationship between the printing parameters and absorption behavior of the fabricated absorbers. The following section describes the bridging and extrude-and-pull methods. Then, the experimental and analytical methods used to characterize their sound absorption performance are described. Finally, we evaluate the efficacy and differences of the two printing methods and study the absorption behavior of the printed samples.

## 2. Methods

### 2.1 Sample Fabrication

#### 2.1.1 Fiber Bridging Method

Fiber bridging involves the continuous extrusion of filament between two points with no underlying support. Schematically presented in Fig. 1(a), this technique relies on extruding the filament at a rate and temperature that allows it to cool and solidify fast enough to bridge a gap without sagging or drooping down to the print bed. While this process is straightforward for small lengths, precise printer settings are necessary to ensure proper bridging over non-trivial distances. A paintbrush with 55 mm long bristles—an approximate aspect ratio of 137—printed using this method is shown in Fig. 1(b). The right-end support can be removed as the final post-processing step to obtain the brush. A schematic representation of the cylindrical test samples printed using this method is shown in Fig. 1(c), where the outer cylindrical shell serves as the bridge support.

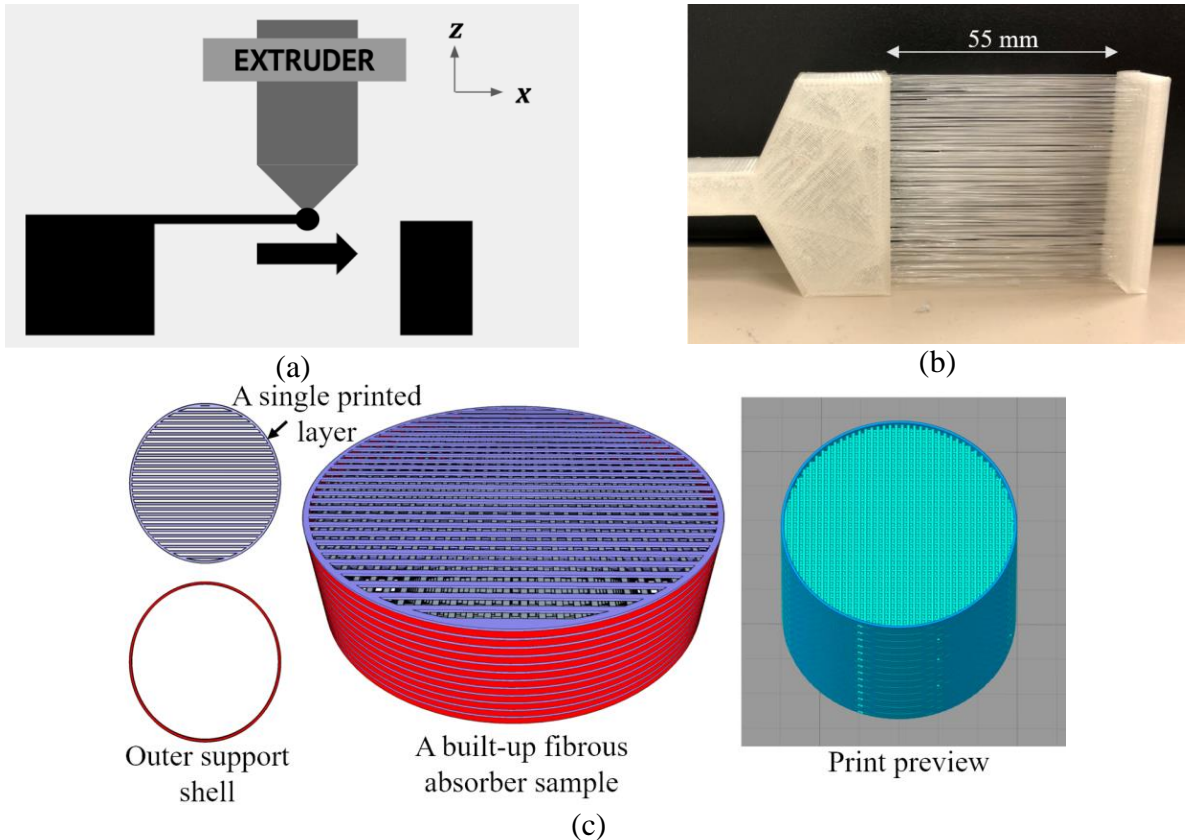

**Figure 1. (a)** Schematic representation of the fiber bridging method, **(b)** a paintbrush printed using this method, and **(c)** schematic representation of cylindrical test samples printed using this method and the print preview.

The fiber bridging process uses the standard 3D printing workflow: model, slice, print. In this study, we use the open-source software OpenSCAD [40] to create the CAD models of cylindrical impedance tube test samples with 30 mm diameter. The fiber density for each sample is controlled by altering the horizontal separation, $\delta_h$, between individual fibers forming each layer and the vertical separation, $\delta_v$, between successive fiber layers, as shown in Fig. 2. Each sample is printed with a constant layer height, $h$, of 0.15 mm. Note that the fiber vertical separation is restricted to multiples of the print layer thickness and limits the achievable fiber vertical separation to 0.15 mm; a lower $\delta_v$ bonds successive fiber layers together and results in effectively no vertical layer separation. All samples are printed using a 0.4 mm printer nozzle width, and the total sample thickness is maintained at 25.4 mm (1-inch). The CAD model information is converted into G-code using the slicing software ideaMaker and then finally printed using a Raise3D N2 Plus FDM printer. Average printing time for each sample created using the fiber bridging method was 90 minutes. All samples studied here are printed using commercially available poly(lactide) (PLA) filaments purchased from Hatchbox3D (www.hatchbox3D.com). While the rheological properties are not explicitly studied here, we expect the properties of filaments used here to be in line with those reported elsewhere [41, 42].

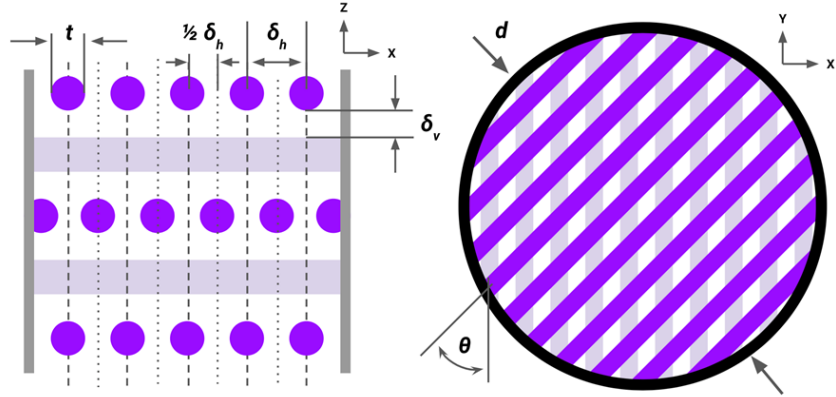

**Figure 2. (Left)** Cross-section and **(right)** top view depiction of fiber parameters.

To understand the absorption behavior of the printed samples, we restrict our focus to three fiber distribution parameters: $\delta_h$, $\delta_v$ and $\theta$–the angle offset between successive layers. Table 1 lists the individual fiber parameters for all samples printed using the fiber bridging method. Note that the samples BR-8, BR-9, and BR-10 were printed with the same distribution parameters but using extruder travel rates of 800 mm/min, 1600 mm/min, and 3200 mm/min, respectively.

**Table 1:** Samples printed using the bridging method and their prescribed parameters.

| Sample Name | Diameter, $d$ (mm) | Angle Offset, $\theta$ (deg) | Horizontal Separation, $\delta_h$ (mm) | Vertical Separation, $\delta_v$ (mm) |
|---|---|---|---|---|
| **BR-1** | 30 | 90 | 0.6 | 0.3 |
| **BR-2** | 30 | 90 | 0.5 | 0.3 |
| **BR-3** | 30 | 90 | 0.8 | 0.3 |
| **BR-4** | 30 | 90 | 0.6 | 0.15 |
| **BR-5** | 30 | 90 | 0.6 | 0.6 |
| **BR-6** | 30 | 45 | 0.6 | 0.3 |
| **BR-7** | 30 | 30 | 0.6 | 0.3 |
| **BR-8** | 30 | 90 | 0.5 | 0.45 |
| **BR-9** | 30 | 90 | 0.5 | 0.45 |
| **BR-10** | 30 | 90 | 0.5 | 0.45 |

### 2.1.2 Extrude and Pull Method

Recently, Laput et al. [37] have proposed an innovative method for 3D printing fibers that exploits the stringing phenomenon commonly observed during the operation of handheld glue guns, where a stringy residue forms when the glue gun is rapidly moved away after extruding the glue. Here, we implement this method to fabricate the second set of fibrous bulk absorbers; we term the method as 'Extrude-and-Pull' since it involves the extrusion of a small amount of material before the print nozzle is rapidly pulled away to generate the thin fibers. Fig. 3(a) shows a schematic representation of this method. We first model the outside cylinder and slice the STL files as before.

The resulting G-code is then modified to add fibers at the desired locations by including commands instructing the printer to extrude a customizable amount of filament at a specific coordinate location on the outer host shell and then rapidly travel to another surface. This method allows the fabrication of hair-like fibers, where the fiber thickness continuously decreases over its length. This can allow further fiber manipulation using post-processing techniques such as curling or braiding. Additionally, unlike the bridging method, it allows the user to control the fiber thickness by changing the extrusion amount and pull speeds. Fig. 3(b) shows a troll couple with 125 mm long hair—an approximate aspect ratio of 312.5—printed using this method.

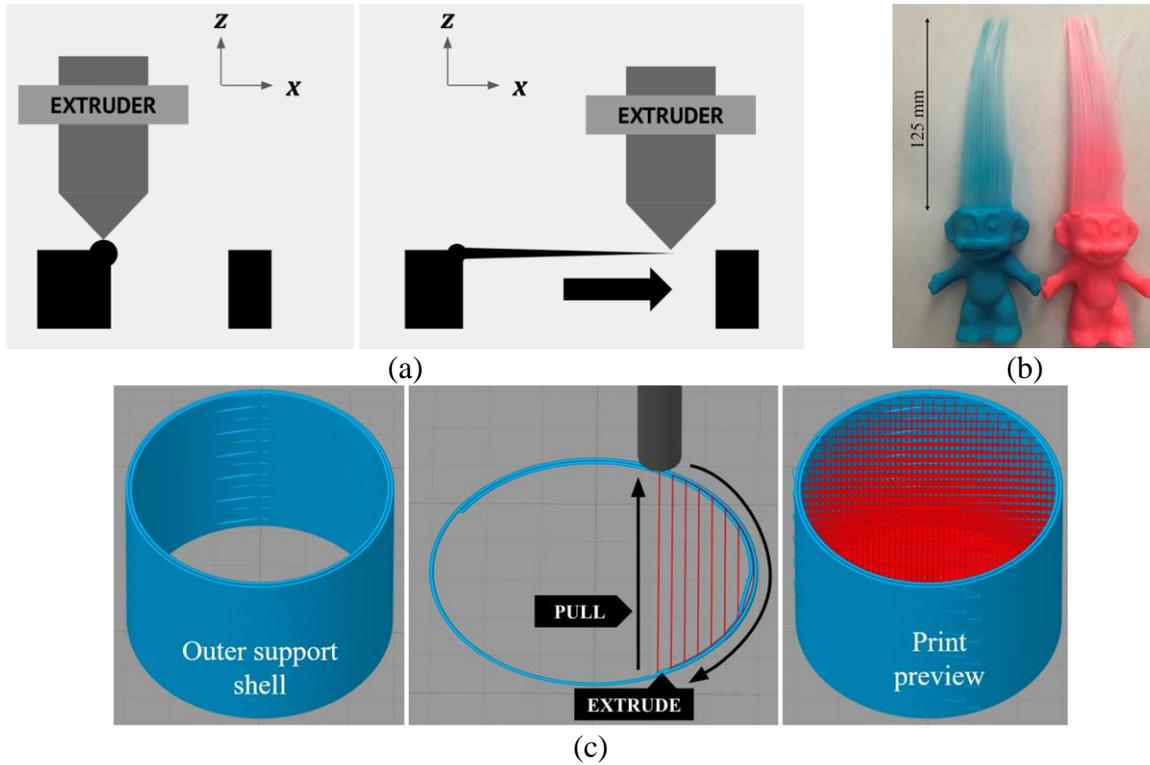

**Figure 3. (a)** Schematic representation of the extrude-and-pull method, **(b)** a troll couple printed using this method, and **(c)** schematic representation of cylindrical test samples printed using this method and the print preview.

We utilized the extrude-and-pull method to fabricate the second set of fibrous absorber samples studied here. The CAD models of the outer shell with the desired sample diameter are sliced using ideaMaker. The generated G-code is then modified to add the fiber information. To avoid manually adding information about each fiber to the G-code—a tedious process that could take multiple hours for each sample—we use a Matlab script to automate this process. The script uses the G-code file for the cylindrical shell as input, determines the coordinates where fibers must be added based on the user-input $\delta_h$ and $\delta_v$ parameters, and then adds the necessary commands instructing the printer to extrude a user-defined amount of filament, $e$, and then pull away at a user-defined speed, $p$, to an opposite point on the shell. This results in fibers of customizable thickness, $t$, spanning the cylinder. The altered G-code file is then fed to the printer for the final printing. The average printing time for each sample created using the extrude-and-pull method was 110 minutes, slightly longer than the bridging method due to the increase printer motion. An excerpt of a modified G-code file is given in Appendix A1.

Table 2 lists the fibrous absorber samples printed using the extrude and pull method. For this case, in addition to the $\delta_h$ and $\delta_v$ parameters, we also alter the extrusion amount ($e$) and pull speed ($p$) to study the effect of fiber thickness on the absorption properties. Since results from the bridging method indicate that $\theta$ does not have any measurable effect on the absorption properties, it was maintained at 90° for all extrude and pull samples.

**Table 2:** Samples printed using the extrude and pull method and their prescribed parameters.

| Sample Name | Diameter, $d$ (mm) | Extrusion, $e$ (Units) | Pull Speed, $p$ (mm/min) | Horizontal Separation, $\delta_h$ (mm) | Vertical Separation, $\delta_v$ (mm) |
|---|---|---|---|---|---|
| EP-1 | 30 | 0.6 | 1600 | 0.8 | 0.6 |
| EP-2 | 30 | 0.6 | 1600 | 0.6 | 0.6 |
| EP-3 | 30 | 0.6 | 1600 | 0.8 | 0.3 |
| EP-4 | 30 | 0.6 | 3200 | 0.8 | 0.6 |
| EP-5 | 30 | 0.6 | 800  | 0.8 | 0.6 |
| EP-6 | 30 | 0.6 | 400  | 0.8 | 0.6 |
| EP-7 | 30 | 0.3 | 1600 | 0.8 | 0.6 |
| EP-8 | 30 | 0.9 | 1600 | 0.8 | 0.6 |

## 2.2 Acoustic Absorption Testing

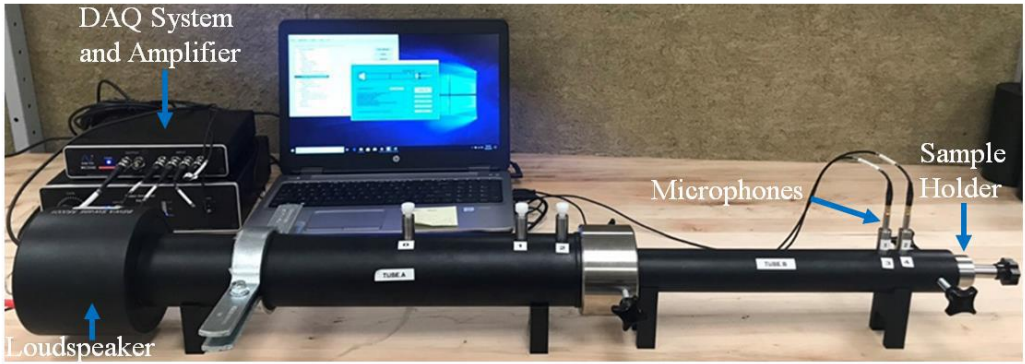

**Figure 4.** The two-microphone normal incidence impedance tube setup used to measure the absorption coefficient of the 3D printed samples.

The sound absorption coefficient of the 3D printed fibrous absorbers is measured using the normal incidence impedance tube measurement method, as prescribed in ASTM E1050-19 [43]. A schematic of the test setup is shown in Fig. 4. The test samples are inserted into the impedance tube using a sample holder with an acoustically rigid backing (no air layer behind the sample). A loudspeaker, driven by a power amplifier, generates white noise with a frequency spectrum of DC–10 kHz. The incident and reflected sound pressure fields are measured using two microphones. All measurements were obtained between a frequency range of 1000 to 6100 Hz, with a reference sound pressure level of 90 dB maintained at the microphone closest to the sample. The microphone transfer function, $H_{12}$—which is the ratio of the pressures measured at microphone 2 and the reference microphone—is measured using Siemens Testlab software. The reflection coefficient, $R$, is then calculated as [43]:

$$R = \frac{H_{12} - e^{-jks}}{e^{jks} - H_{12}} e^{j\,2k(l+s)} \tag{1}$$

where $k$ is the wavenumber in the fluid media (air), $l$ and $s$ are the distance between the front surface of the sample to its nearest microphone, and microphone spacing, respectively. Consequently, the sound absorption coefficient, $\alpha$, and the normalized surface impedance, $z$, are then calculated as:

$$\alpha = 1 - |R|^2 \tag{2}$$

$$\frac{z}{\rho_o c} = \frac{1+R}{1-R} \tag{3}$$

where $R^*$ is the complex conjugate of the reflection coefficient, $\rho_o$ is the density of air (1.21 kg/m³), and $c$ is the speed of sound in air (343 m/s). For consistency, all samples were tested at a 90 dB sound pressure level, as measured at the microphone location closest to the sample.

## 2.3 Modeling: Inverse Characterization

Sound absorption in fibrous porous structures is attributable to the frequency-dependent visco-inertial and thermal losses that occur as the incident sound waves travel through the fibrous solid skeleton. While it is difficult to develop an exact analytical model of this complex energy conversion process, various semi-empirical models have been proposed over the years. A comprehensive discussion of these models and their historical developments is available in Ref. [6]. Here, we model the acoustical behavior of the 3D printed fibrous absorbers using the Johnson-Champoux-Allard (JCA) model. In the JCA model [38, 39], the solid frame is modeled as a rigid structure—a suitable assumption when the frame motion is negligible as compared to the motion of the fluid phase—and the acoustical properties are derived as functions of five microstructure-dependent transport parameters [6]: open porosity, static air flow resistivity, tortuosity, viscous characteristic length, and thermal characteristic length. These five properties may either be obtained directly through experimental measurements [6], calculated using various numerical approaches [44], or derived using an inverse characterization method using impedance tube measurements [45-47].

Here, we adopt the inverse method to characterize the transport properties of the 3D printed fibrous absorbers and gain further insights into their behavioral differences. The inverse characterization method is a curve fitting approach where the five transport parameters are independently varied until the model predictions match the experimentally measured results. The best-fit parameters are obtained by minimizing a merit function designed to quantify the difference between the predictions and the measurements. In this study, we use the commercial software FOAM-X to perform the inverse characterization. The sound absorption data obtained from impedance tube measurements was used as the experimental reference data, and the five transport parameters were varied until the merit function is minimized. Further details about the inverse characterization method can be obtained in Ref. [45-47].

## 3. Results and Discussion

### 3.1 Sample Print Quality

Representative examples of printed fibrous structures and observed print quality issues are shown in Fig. 5. It was noticed that, on average, some fiber parameter choices affect the overall print quality more than others and affect the two methods differently. For instance, increased fiber density using the bridging method sometimes resulted in fiber bending (Fig. 5(a)), likely because of the heat emanated from the printer's extruder warping adjacent fibers. Meanwhile, decreased $\delta_h$ and increased $\delta_v$ led to the best results for the extrude-and-pull method. For the bridging method, extruder travel rates of 800 mm/min and below resulted in samples of noticeably poor quality, such as connected adjacent fibers (Fig. 5(b)), broken fibers, excessive warping, and weak fiber-shell bonding (Fig. 5(c)); the extrude-and-pull method resulted in samples of acceptable quality for rates of 400 mm/min and above. During the extrude-and-pull method, too small extrusion amounts (less than 0.3 mm) result in unformed, under-extruded fibers (Fig. 5(d)), while over extrusion (greater than 1.5 mm) leaves globs of filament behind each fiber. The angle offset and sample diameter choices had no noticeable impact on print quality. Representative examples of fibrous samples deemed unacceptable due any of the above quality issues is shown in Fig. 5(e), while an example of a high-quality samples is shown in Fig. 5(f). Overall, both methods provided high print repeatability and limited number of failures occurred during this study.

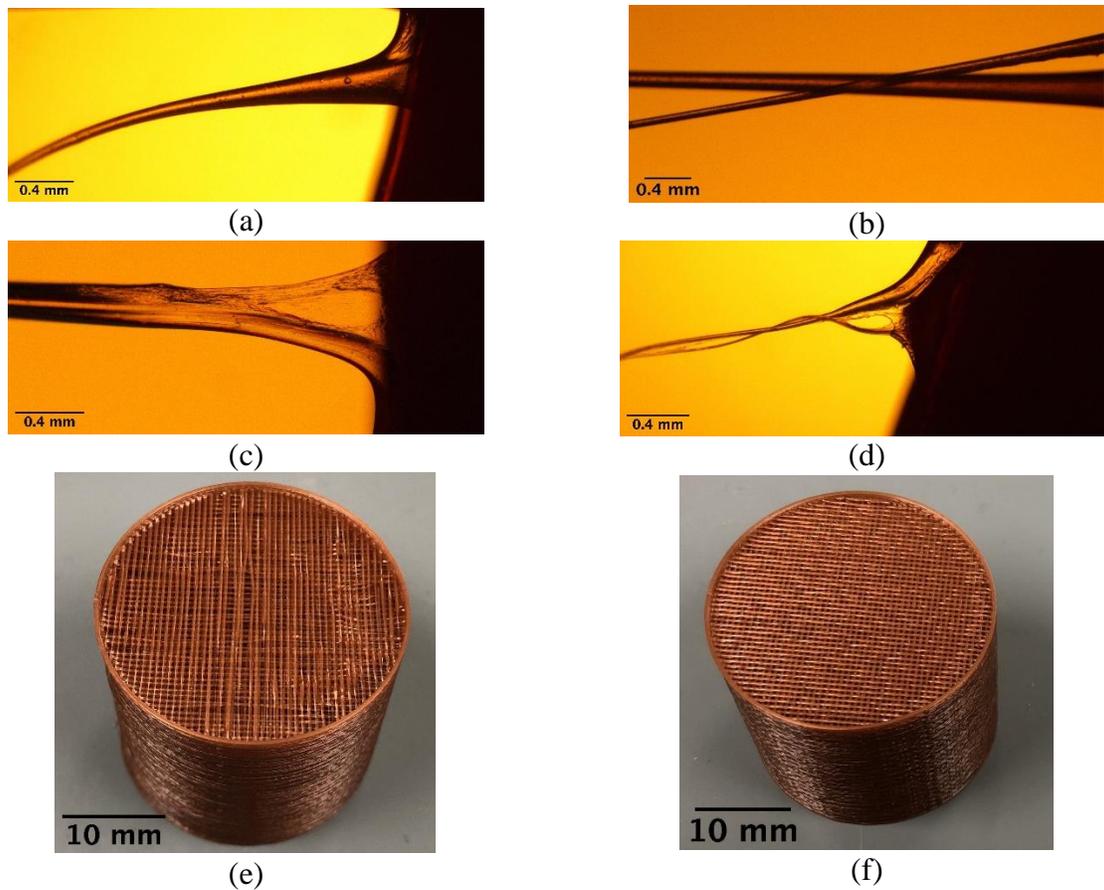

**Figure 5.** Representative examples of observed printing issues: **(a)** excessive fiber bending; **(b)** connected adjacent fibers; **(c)** weak connection of fiber with outer wall; **(d)** under-extrusion of filament. Representative example of **(e)** rejected samples; **(f)** accepted samples.

The effect of the primary printing parameters on the fiber thickness was studied by varying individual parameter—extruder travel rate for the bridging method, and extrusion amount and pull rate for the extrude-and-pull method—and printing samples with a single layer of fibers with lengths varying from 10 mm to 40 mm. The fiber thickness variation along its length was measured using an optical microscope (AmScope 40X-800X Trinocular Dual-illumination Metallurgical Microscope with Polarization) with a dedicated 18-megapixel camera. The thickness measurements were done every 5 mm along the fiber lengths and are shown in Fig. 6. For brevity, only results for the 40 mm long fibers are presented here; the trends for the other lengths were the same.

Fig. 6(a) shows the effect of extruder travel rate on the thickness of fibers printed using the bridging method. For all travel rates, the fiber thickness initially decreases, achieving a minimum thickness of about 0.1 mm around 5 mm from the fiber starting point, before gradually increasing over its length, and converging to a final thickness equal to the extruder nozzle diameter of 0.4 mm. The initial reduction of fiber diameter can be attributed to the delay in filament extrusion as the extruder head starts traveling towards the final fiber coordinate. Overall, a faster extruder travel rate produces fibers with lower thicknesses.

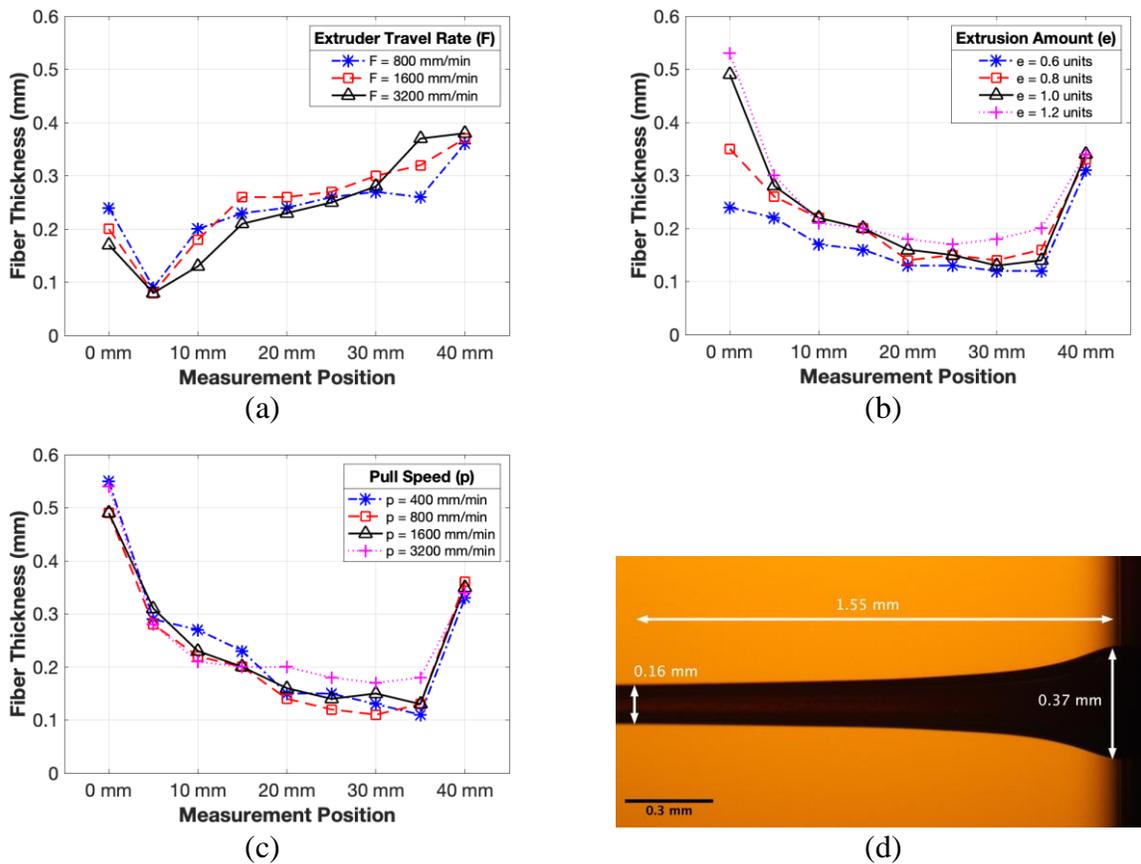

**Figure 6.** Variation of fiber thickness along the printed fiber length. The parameters varied are **(a)** extruder travel rate for the bridging method, and **(b)** extrusion amount and **(c)** pull speed for extrude-and-pull method. **(d)** A representative image showing the abrupt increase in fiber thickness at the fiber end.

Figs. 6(b) and 6(c) show the effect of extrusion amount and pull speed on fiber thickness variation during the extrude-and-pull method. The fiber thickness characteristics are noticeably different from those printed using the bridging method. For the extrude-and-pull method, the fibers are thickest near the extrusion location. As the extruder rapidly pulls away from the extrusion point, the fiber thickness gradually decreases to less than half of the initial fiber thickness before abruptly increasing during the final 5 mm of the fiber length and converging to the nozzle diameter. This abrupt increase in fiber thickness, shown in Fig. 6(d), occurs because of the inadvertent oozing out of filament as the extruder reaches the fiber end location and prepares itself to print the next layer segment. This oozing may be controlled by including a coasting command or increasing the filament retraction amount. In this study, no coasting or wiping commands were included in the G-code and effects of including these steps will be investigated in future studies. As expected, the initial fiber thickness depends heavily on the extrusion amount and a larger extrusion amount results in thicker fibers over their entire length. Interestingly, though the pull speed does not affect the initial fiber thickness, faster pull speeds produce relatively thicker fibers overall.

## 3.2 Sound Absorption Behavior

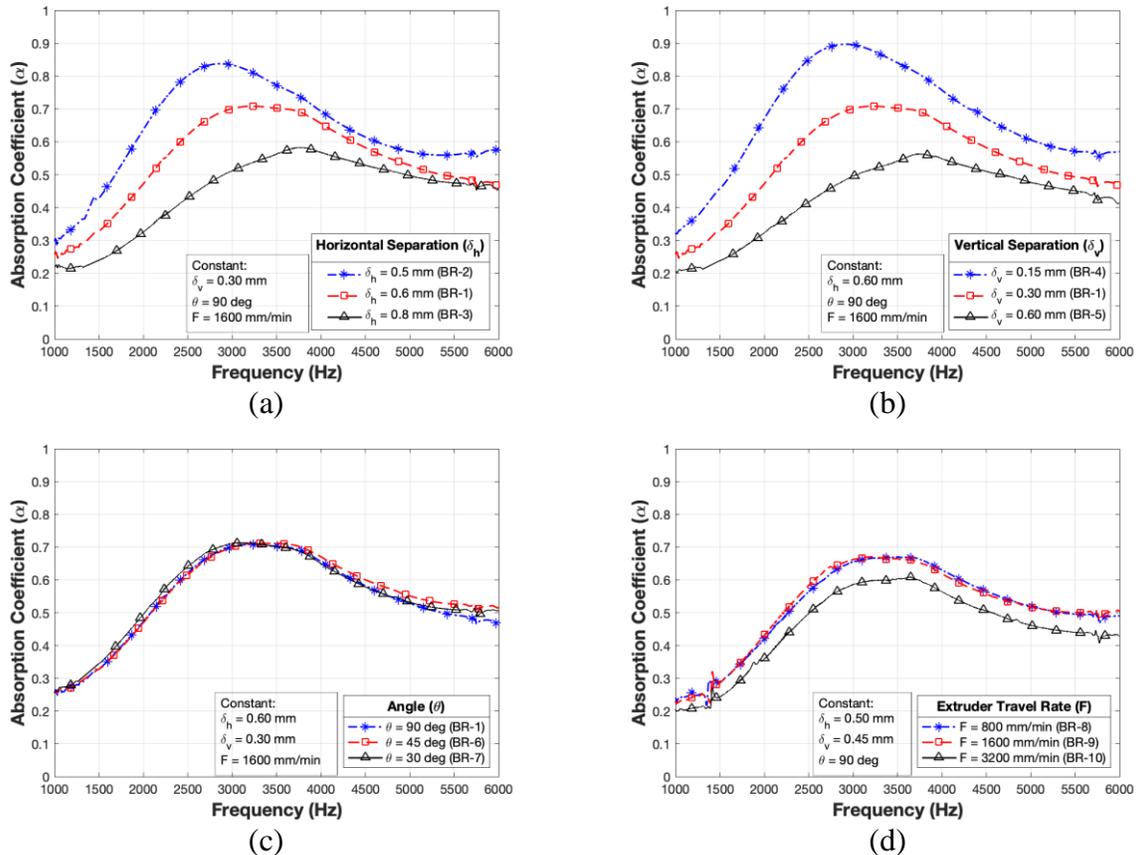

**Figure 7.** Effect of individual printing parameters on the sound absorption behavior of samples printed using the bridging method. The varying parameters are: **(a)** $\delta_h$, **(b)** $\delta_v$, **(c)** $\theta$, and **(d)** extruder travel rate.

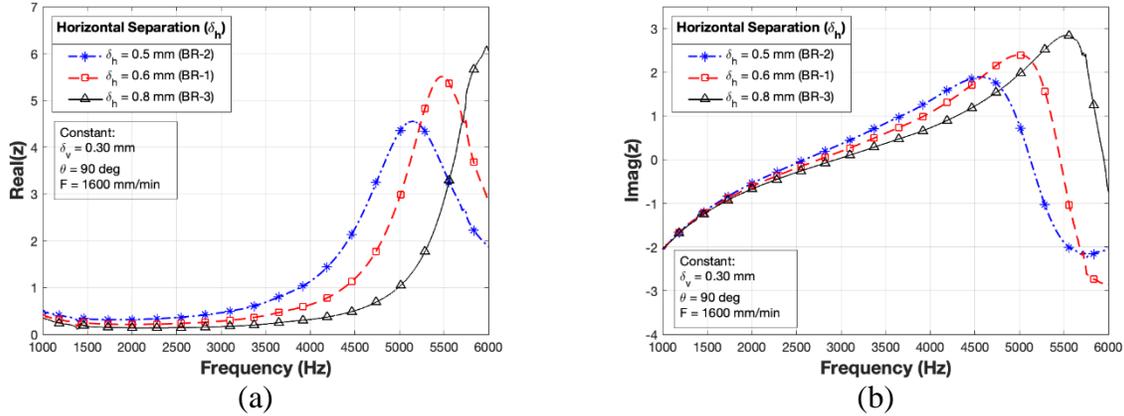

**Figure 8.** Effect of varying $\delta_h$ on the **(a)** real part (or resistance) and **(b)** imaginary part (or reactance) of the normalized surface impedance (z) of samples printed using the bridging method.

The effects of various user-defined parameters and extruder travel rates on the sound absorption behavior of sample printed using the bridging method are compared in Fig. 7. Representative normalized surface impedance curves are shown in Fig. 8. For brevity, only the impedances obtained for the samples with varying $\delta_h$ are shown; impedances for all other cases behave similarly. The overall absorption behavior of all samples follows a trend commonly observed for open-celled porous structures [6]: the absorption gradually increases with increasing frequency and reaches a peak absorption value, followed by a trough with reduced absorption performance. As observed from the impedance curves, the absorption is primarily resistance-driven, with the peak absorption value corresponding to the depth resonance frequency [6]. Fig. 7(a) shows the effect of varying horizontal separation ($\delta_h$) while maintaining a constant vertical separation ($\delta_v$) and inter-layer angle ($\theta$). As $\delta_h$ is decreased from 0.8 mm to 0.5 mm, the sound absorption at all frequencies increases and the absorption peak shifts to a lower frequency. A similar trend is observed when $\delta_v$ is varied while maintaining constant $\delta_h$ and $\theta$ values (Fig. 7(b)). This increase in absorption because of reducing $\delta_h$ or $\delta_v$ results from the increased fiber density (i.e., smaller pore sizes) that causes increased low-frequency visco-inertial losses and high-frequency visco-thermal losses. This effect is verified by observing Fig. 8, where the resistance values increase with decreasing $\delta_h$. Changing the inter-layer angle does not result in any appreciable change in absorption (Fig. 7(c)). The effect of varying extruder travel rate is shown in Fig. 7(d). The absorption coefficients of samples printed using a 3200 mm/min travel rate are noticeably lower than those printed using 800 mm/min and 1600 mm/min travel rates. This can be explained by referring to the effect of travel rates on the fiber thickness (Fig. 6(a)): faster rates produce thinner fibers, thus resulting in effectively larger pore sizes and reduced sound energy dissipation.

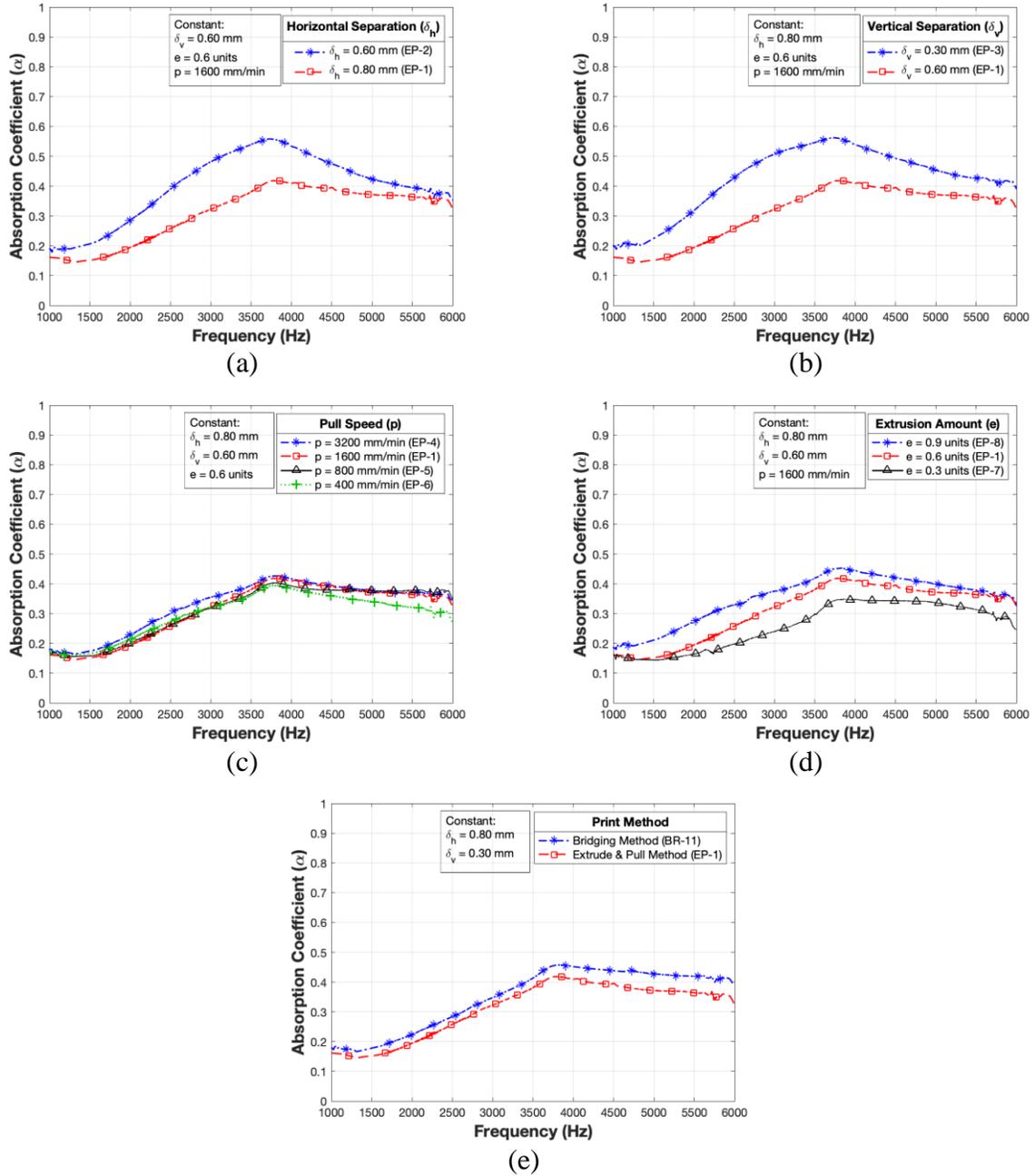

**Figure 9.** Effect of individual printing parameters on the sound absorption behavior of samples printed using the extrude-and-pull method. The varying parameters are: **(a)** $\delta_h$, **(b)** $\delta_v$, **(c)** pull speed ($p$), and **(d)** extrusion amount ($e$). **(e)** Comparison of absorption coefficients of samples printed using the bridging and extrude-and-pull methods.

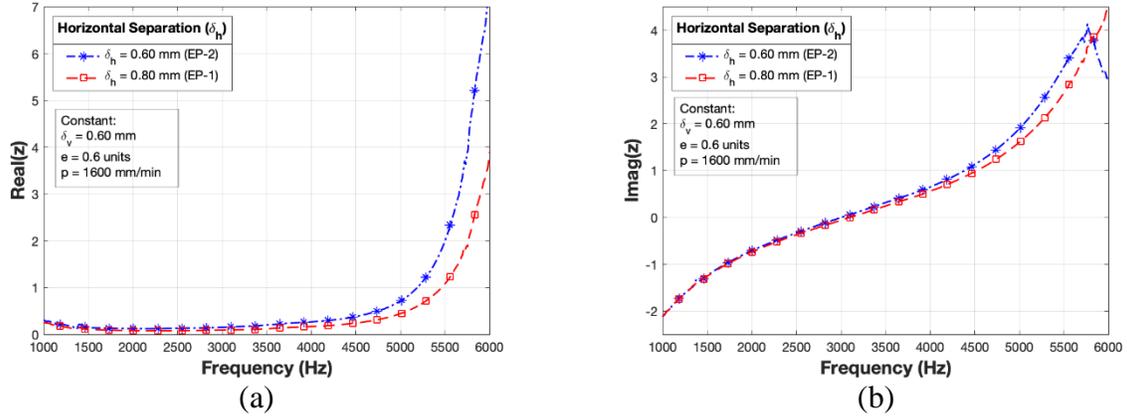

**Figure 10.** Effect of varying $\delta_h$ on the **(a)** real part (or resistance) and **(b)** imaginary part (or reactance) of the normalized surface impedance (z) of samples printed using the bridging method.

Absorption test results for samples printed using the extrude and pull method are shown in Fig. 9. As seen for the bridging method, increased fiber density through lower $\delta_h$ or $\delta_v$ values result in increased sound absorption (Figs. 9(a) and (b)). The impedance curves, shown in Fig. 10 for the case of varying $\delta_h$, are similar to those observed for the bridging method. For a constant extrusion amount ($e$), increasing the pull speed ($p$) results in a slightly higher sound absorption value (Fig. 9(c)). This occurs because of the increased thickness of the fibers printed using higher pull speeds—as observed in Fig. 9((b))—which results in an increase in the effective fiber density. Interestingly, the increase in energy dissipation is comparatively more pronounced at lower frequencies, suggesting that the generated fiber thickness variations affect the static flow resistivity more than the other transport parameters. Similarly, an increase in filament extrusion amount ($e$) while keeping all other parameters constant (Fig. 9(d)) results in higher sound absorption because of the thicker fibers and increased effective fiber density. A comparison of sound absorption behavior of samples with the same $\delta_h$ and $\delta_v$ values printed using the two different methods is shown in Fig. 9(e). While the overall behavior of the two samples is quite similar, the sample printed using the bridging method provides consistently higher absorption than the extrude-and-pull sample over the measured frequency range. This is consistent with results from the fiber thickness analysis; the average thickness of fibers printed using the bridging method is greater than those printed using the extrude-and-pull method, which results in smaller effective pore sizes and greater sound energy dissipation.

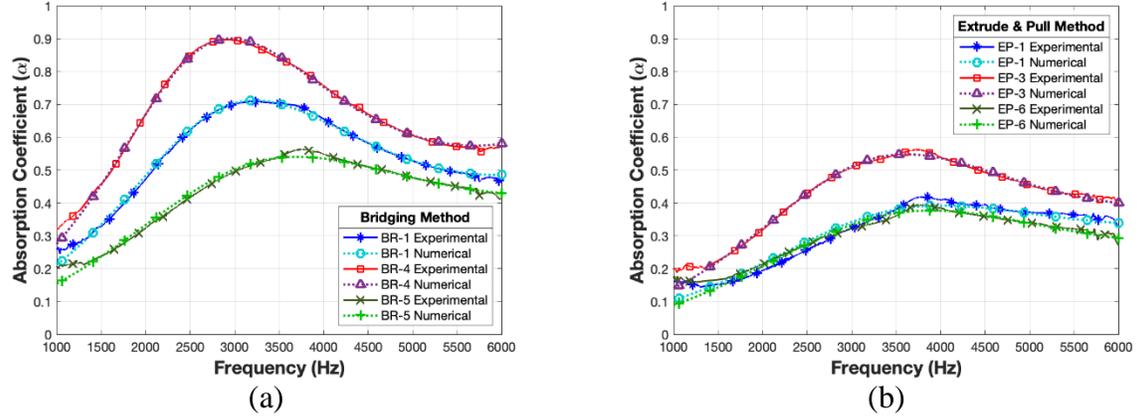

**Figure 11.** Comparison of experimental measurements and numerical predictions obtained using the inverse characterization method and JCA model for samples printed using the **(a)** bridging method and **(b)** extrude-and-pull method.

The actual absorption values for the different samples are a function of their individual transport properties. The transport properties for all tested samples were obtained using the inverse characterization method and the samples are modeled as rigid absorbers using the five-parameter JCA formulation. Of these parameters, the porosity is a geometrical parameter and is calculated by dividing the mass of the empty volume within the porous structure by the total mass of an equivalent solid volume. Here, the mass of the empty volume is calculated by weighing each sample and dividing it by the typical density of a PLA filament—1.24 g/cm$^3$. The calculated porosity, given in Tables 3 and 4, is then used as a known parameter during the curve fitting to help increase the algorithms accuracy. A comparison of the analytically predicted absorption curves with the measured values is shown in Fig. 11. For clarity, results from only three samples for each method are plotted; similar results were obtained for all other cases. While some deviations can be seen at lower frequencies (below 1500 Hz), the measured and predicted results show an excellent match over the frequency range of interest. The deviations may be attributed to the lower-quality low-frequency data typically obtained using the impedance tube measurement method. This issue will be further investigated in the future using larger samples sizes that allow more accurate low-frequency measurements. For this study, we restrict our focus to the 1500-6000 Hz frequency range.

**Table 3.** Transport parameters obtained using the inverse characterization method for the bridging samples.

| Sample Name | Porosity | Flow Resistivity (N.s.m$^{-4}$) | Tortuosity | Viscous Characteristic Length (μm) | Thermal Characteristic Length (μm) |
|---|---|---|---|---|---|
| BR-1 | 0.757 | 6179 | 1.189 | 280 | 280 |
| BR-2 | 0.719 | 6082 | 1.246 | 176.1 | 191.3 |
| BR-3 | 0.798 | 2562 | 1.021 | 304.4 | 304.4 |
| BR-4 | 0.659 | 9875 | 1.164 | 152.3 | 199.7 |
| BR-5 | 0.835 | 3868 | 1.087 | 397 | 397 |
| BR-6 | 0.757 | 5080 | 1.12 | 240.5 | 240.9 |

| | | | | |
|---|---|---|---|---|
| BR-7 | 0.758 | 4559 | 1.186 | 235.6 | 235.6 |
| BR-8 | 0.79 | 4719 | 1.152 | 269.6 | 301.4 |
| BR-9 | 0.786 | 2869 | 1.168 | 237.7 | 271.6 |
| BR-10 | 0.789 | 4220 | 1.168 | 350.3 | 350.3 |

The predicted transport properties—presented in Tables 3 and 4—provide further insight and explanation for the observed absorption trends. For the bridging method, a reduction in the vertical and horizontal fiber separation ($\delta_h$ variation: samples BR-1, -2, -3; $\delta_v$ variation: samples BR-1, -4, -5) causes an increase in air flow resistivity and tortuosity while decreasing the viscous and thermal characteristic lengths. The air-flow resistivity and tortuosity govern the low-frequency behavior which is dominated by the visco-inertial response of the structure [6]. Thus, an increase in these values corresponds to the higher absorption values obtained in the lower-frequency regions. The shift of the peak absorption values to a lower frequency is primarily a result of the increase in the tortuosity of the samples, indicating that the reduced fiber separation results in a more tortuous path for the incident sound waves. The high-frequency response is dominated by the visco-thermal response of the structure and primarily governed by the viscous and thermal characteristic lengths [6]. Reduced characteristic lengths result in the higher absorption observed at frequencies higher than the peak absorption frequency. For samples with varying inter-layer angles ($\theta$ variation: samples BR-1, -6, -7), no significant variation of the values is observed between the three samples, which is reflected in the similar absorption behavior observed experimentally. Interestingly, variation in the extruder travel rate (samples BR-8, -9, -10) results in samples with similar tortuosity and characteristic lengths and the difference in absorption behavior results from the reduced resistivity of samples with thinner fibers printed at higher extruder travel rates. A porosity of 0.85 is predicted for all samples printed using the bridging method.

**Table 4.** Transport parameters obtained using the inverse characterization method for the extrude-and-pull samples.

| Sample Name | Porosity | Flow Resistivity (N.s.m$^{-4}$) | Tortuosity | Viscous Characteristic Length (μm) | Thermal Characteristic Length (μm) |
|---|---|---|---|---|---|
| EP-1 | 0.82 | 1048 | 1 | 468.8 | 478.4 |
| EP-2 | 0.79 | 3923 | 1.074 | 393.8 | 393.8 |
| EP-3 | 0.77 | 4033 | 1.061 | 456.2 | 456.2 |
| EP-4 | 0.82 | 1721 | 1.004 | 464.9 | 464.9 |
| EP-5 | 0.82 | 1100 | 1 | 436.1 | 609.2 |
| EP-6 | 0.82 | 2722 | 1.025 | 773.7 | 773.7 |
| EP-7 | 0.86 | 1528 | 1 | 682.6 | 682.6 |
| EP-8 | 0.79 | 2609 | 1 | 474.5 | 474.5 |

The behavior of samples printed using the extrude-and-pull method is noticeably different. <span style="color:red">In general, samples printed using this method higher porosities than the bridging samples.</span> For the samples with varying fiber separations ($\delta_h$ variation: samples EP-1, -2; $\delta_v$ variation: samples EP-1, -3), the tortuosity values remain similar though the flow resistivity and characteristic length values are significantly different. This is reflected in the absorption curves; while the absorption curves shift to higher values over the entire frequency range for samples with smaller $\delta_h$ and $\delta_v$, the peak absorption occurs around the same frequency for all cases. The variation in flow resistivity and characteristic lengths can be explained by the variation in thickness over the length of the fibers printed using this method. As seen in Figs. 6(b-c), the individual fiber thickness varies between a maximum value of approximately 0.5 mm at the fiber end and a minimum value of approximately 0.15 mm towards the central portion of the fiber. The effect of thickness variations is also apparent in samples with varying pull speed (samples EP-1, -4, -5, -6) and varying extrusion amount (samples EP-1, -7, -8)—despite maintaining the same geometrical parameters, the absorption amounts vary significantly depending on the pull speed and extrusion amount. Further, as opposed to the bridging case, the extrude-and-pull samples show a wider spread in porosity, with samples printed at low pull speeds (EP-5 and EP-6) and low extrusion amount (EP-7) showing the most variation. It should be noted that the resistivity values for samples EP-5 and EP-7 coincide with the lower bound prescribed for accurate characterization using the optimizing algorithms implemented in FOAM-X and may thus not be accurate. These uncertainties correspond to the poor sample quality—and hence poor impedance measurements—for samples printed with very low pull speed and extrusion amounts. However, overall, the print parameter choice has a significant effect on the absorption behavior of the resulting samples. As seen in Fig. 9(e), the difference in the samples printed using the two different methods is also clear in their respective transport properties—the fibers printed using the extrude-and-pull method are thinner than those printed using the bridging method and result in lower visco-inertial and thermal losses and provide lower absorption over the entire frequency range.

**Conclusion**
Two methods for additively manufacturing fibrous sound absorbers are presented. The first method—fiber bridging—involves the continuous extrusion of filament between two points with no underlying support. The second method—extrude-and-pull—requires the extrusion of a small amount of heated filament before the print nozzle is rapidly pulled away to generate the thin fibers. Microscope analysis shows that both methods result in fibers with distinct characteristics: fiber bridging results in relatively thicker fibers with the average thickness inherently linked to the nozzle diameter; extrude-and-pull results in fibers with lower average thicknesses, where the thickness is a function of extrusion amount and extruder pull speed. The sound absorption characteristics of samples printed using the two methods are studied using a two-microphone normal incidence test method. It is found that the absorption characteristics of the printed fibrous absorbers depend on the printing parameters—the extruder travel rate for the bridging method, and extrusion amount and pull speed for the extrude-and-pull method—and on the fiber density parameters: the fiber vertical ($\delta_v$) and horizontal ($\delta_h$) separations. Thus, the absorption behavior of such samples can be tailored by appropriately tuning these parameters. An inverse characterization method coupled with the JCA analytical model is used to study the effect of printing parameters on the acoustical transport properties of the samples printed using the two methods. It is observed that the lower individual fiber thickness of extrude-and-pull samples results in comparatively lower visco-inertial and thermal losses and results in their lower absorption performance over the studied frequency range.

Here, 3D printing of fibers was discussed within the context of sound absorption devices. While the overall absorption obtained here is lower than that obtained using traditional fibrous materials, the addition of fibers to other 3D printed porous structural geometries can be used to help improve their noise reduction potential and enable multifunctionality. Further, the printing of fibrous structures can be applicable to a wide range of engineering applications and problems. An obvious application of the presented methods could be the printing of household utility brush tools—paintbrushes, toothbrushes, sweepers, etc. The extrude-and-pull method can be leveraged for flight tests and wind tunnel based aerodynamic studies reliant on the use of tufts for flow visualization [48]. Currently used tufts are manually secured to the test model and must be cut to specific lengths and weight for unaltered flow visualization. The proposed extrude-and-pull method can be easily incorporated into a 3D printing routine to print flight models with incorporated surface tufts as flow indicators. Such fibers, if printed using conductive filaments, may also help improve 3D printed health-monitoring [49] and energy storage devices [50]. Addition of small fibers to base geometries can help manipulate surface textures [51] and enable acoustic or flow based sensors and actuators [52]. Thus, the presented work helps lay the foundation for future studies towards solving a wide array of engineering challenges requiring fibrous structures integrated within structural geometries.

**Acknowledgement**

This work was supported by a NASA EPSCoR Cooperative Agreement Notice (Grant Number: 80NSSC19M0153). William Johnston would also like to acknowledge the support provided by the Robert W. Young Award for Undergraduate Student Research in Acoustics presented by the Acoustical Society of America.

**Appendix**

Below is an excerpt from the G-code file used to print Sample EP-1 ($e = 0.6$ mm, $\delta_h = 0.8$ mm, $\delta_v = 0.6$ mm, $p = 1600$ mm/min) showing the process of forming two fibers within layer 19. Using the extrude and pull method, an empty cylindrical shell is sliced into G-code using ideaMaker, then a Matlab program is used to edit the G-code to add customizable fibers. On the inside of the cylindrical shell, the printer extrudes an amount of filament, $e$, then travels at speed $p$ to the opposite end of the shell to terminate the fiber. Then, the extruder travels along the circumference of the shell before returning to the start of the fiber. Subsequent fibers are printed $\delta_h$ apart.

```
[…]
; LAYER:19
; Z:4.970
; HEIGHT:0.150
; TYPE: WALL-OUTER
G1 X139.135 Y156.981 E829.2366
G1 X139.029 Y156.712 E829.2590
G1 X138.927 Y156.441 E829.2814
G1 X138.831 Y156.168 E829.3038
G1 X138.440 Y155.793 E829.3640
; First Fiber Formation:
; Extrude 0.6 mm:
G1 F728 X138.393 Y155.600 E829.964
; Pull away at 1600 mm/min:
G1 F1600 X166.651 Y155.600
; Travel back to fiber start:
G1 X166.586 Y155.882
G1 X166.516 Y156.163
```

```
G1 X166.440 Y156.442
G1 X166.358 Y156.719
G1 X166.271 Y156.995
G1 X166.178 Y157.269
[…]
G1 X138.643 Y155.575
G1 X138.564 Y155.297
G1 X138.491 Y155.017
G1 X138.324 Y154.936
G1 X138.262 Y154.853
```
**; Second Fiber Formation (0.8 mm away from previous fiber):**
**; Extrude 0.6 mm:**
```
G1 F728 X138.241 Y154.800 E830.564
```
; **Pull away at 1600 mm/min:**
```
G1 F1600 X166.803 Y154.800
```
; **Travel back to fiber start:**
```
G1 X166.754 Y155.085
G1 X166.700 Y155.369
G1 X166.639 Y155.652
G1 X166.573 Y155.934
G1 X166.502 Y156.214
G1 X166.425 Y156.493
G1 X166.342 Y156.770
[…]
```